\documentclass[12pt]{article}
\usepackage{amsfonts}
\usepackage{amssymb}
\DeclareMathSymbol{\rtimes}{\mathbin}{AMSb}{"6F}

\def\bel{\begin{equation}\label}
\newcounter{lit}
\newenvironment{lit}
               { \stepcounter{equation}
             \setcounter{lit}{\value{equation}}
             \setcounter{equation}{0}
             
                }
           { \setcounter{equation}{\value{lit}}
             
                }

\begin{document}

\title{Relation between quantum $\kappa$-Poincar\'{e} framework
 and Doubly Special Relativity }

\author{Jerzy Lukierski\\ Institute of Theoretical Physics,
Wroc{\l}aw University,\\
50-204 Wroc{\l}aw, pl. M. Borna 9, Poland}
\date{}

\maketitle

\begin{abstract}
We describe firstly the basic features of quantum
$\kappa$-Poincar\'{e} symmetries  with their Hopf algebra
structure. The quantum $\kappa$-Poincar\'{e} framework in any
basis relates rigidly the quantum $\kappa$-Poincar\'{e} algebra
 with quantum $\kappa$-Poincar\'{e}
  group,  noncommutative space-time and $\kappa$-deformed phase space.
 Further we present the approach of Doubly Special
Relativity (DSR) theories, which introduce
 (in the version DSR1)  kinematically  the frame -
independent fundamental mass parameter as described by maximal
three-momentum
$|\overrightarrow{p}|=\kappa c$.
 We argue  why the  DSR theories  in one-particle sector
  can be treated as the
 part of quantum $\kappa$-Poincar\'{e} framework. The
DSR formulation has been  extended to multiparticle states
 either in a way leading to nonlinear
description of classical relativistic symmetries, or
 providing  the identification of DSR approach with full quantum
$\kappa$-Poincar\'{e} framework.
\end{abstract}

\section{Introduction}
The aim of this paper is to present the relation between the
formalism of quantum $\kappa$-Poincar\'{e} symmetries
($\kappa$-Poincar\'{e} quantum algebra  and $\kappa$-deformed
 energy-momentum
dispersion relation [1--8], $\kappa$-Poincar\'{e} quantum group and
$\kappa$-Minkowski space [9,4--6],
 $\kappa$-deformed phase space [4--6,10])
 and
recently developed Doubly Special Relativity (DSR) theories [11--19].

Quantum $\kappa$-deformed Poincar\'{e} symmetries are described by
quite rigid scheme of Hopf algebras which generalize the notion of
classical symmetries to the case of noncommuting transformation parameters.
Such a scheme describes simultaneously the deformation of
classical Poincar\'{e} algebra $P^{3;1}$

\begin{lit}
\begin{eqnarray}
\label{proclu1a}
{}[M^{(0)}_{\mu\nu},M^{(0)}_{\rho\tau}]
&=&i(g_{\mu\rho}M^{(0)}_{\nu\tau}
+ g_{\nu\tau}M^{(0)}_{\mu\rho} - g_{\mu\tau}M^{(0)}_{\nu\rho} -
g_{\nu\rho}M^{(0)}_{\mu\tau})
\\
\label{proclu1b}{}[M^{(0)}_{\mu\nu},P^{(0)}_\rho]
&=&i(g_{\mu\rho} P^{(0)}_\nu
-
g_{\nu\rho}P^{(0)}_\mu)
\\
\label{proclu1c}
[P^{(0)}_\mu,P^{(0)}_\nu]
&=&0
\end{eqnarray}
\end{lit}
\hskip-12pt
 as well as
 the deformation of classical Poincar\'{e} group
${\cal P}^{3;1}=O(3,1)\rtimes T_4$
 to quantum one, with  noncommuting translation and Lorentz
rotation parameters [9,6]. Using the duality
of Hopf algebras (see e.g. [20]) one can show that the deformation of quantum
Poincar\'{e} group (e.g. $\kappa$-Minkowski space-time described by the
translation sector $T_4$) follows uniquely from the Hopf-algebraic structure
of quantum $\kappa$-Poincar\'{e} algebra.

Doubly Special
Relativity (DSR) theories guided by some theoretical challenges in
astrophysics and quantum gravity
 (see e.g. [21--24]) derive their framework from the
 explicite form of
$\kappa$-deformed mass-shell condition originally postulated in
DSR1 theories as follows (see e.g. [15])
\begin{equation}\label{proclu2}
C^{\kappa}_2 (\overrightarrow{p}^{\, 2}, p_{0})= \left( 2\kappa \,
\sin \frac{P_0}{2\kappa} \right)^2 -
\overrightarrow{P}^{\, 2}\, e^{\frac{P_0}{\kappa}} = M^2\, ,
\end{equation}
with the deformation parameter $\kappa$ identified with
the Planck mass $M_{\mu\nu}$.
It appears that (\ref{proclu2}) is identical with particular form of
$\kappa$-deformed mass Casimir for  quantum
$\kappa$-Poincar\'{e} algebra\footnote{For quantum $\kappa$-Poincar\'{e}
algebra in DSR papers there is  used the new name of
$\kappa$-Poincar\'{e}-Hopf algebra.} in bicrossproduct basis [4,5,25].

We shall present firstly in Sect. 2 the basic features of the
Hopf-algebraic framework of quantum $\kappa$-Poincar\'{e}
symmetries, in particular we will describe the  Hopf-algebraic approach
 the $\kappa$-deformed Lorentz transformations
of fourmomenta
 and will point out the arbitrariness of the choice
of basic generators defining quantum $\kappa$-Poincar\'{e}
algebra.
Further the interpretation (see [8]) of the classical Lie algebra basis
for quantum $\kappa$-deformed Poincar\'{e}  algebra will be
 given.
 In Sect. 3 we shall consider the $\kappa$-deformed   Lorentz
transformation in DSR1 theories and show that they are a part of
the framework of quantum $\kappa$-Poincar\'{e} algebra.
We shall point out the absence of definite coproduct rules in
DSR framework, and recall two
alternative  choices proposed in [17].
 Further
we recall that one can introduce three classes of DSR theories [26],
with bounded $|\overrightarrow{p}|$ (DSR1), bounded $E$ (DSR3) and
both variables $|\overrightarrow{p}|$ and $ E$ bounded (DSR2)
 with the first model in DSR2 class provided by Magueijo and
  Smolin [14]. In Sect. 4 we present final remarks.

\section{Hopf-algebraic structure of quantum $\kappa$-Poincar\'{e}
symmetries}

The quantum $\kappa$-Poincar\'{e} algebra ${ U}_\kappa \,P^{3;1}_{\kappa}$,
introduced in 1991   [1--3] has been rewritten in 1994 [4--6] in so-called
bicrossproduct  basis as follows:

a) algebraic sector ($M_{\mu\nu}=(M_i, N_i$); $P_\mu =
(P_i, P_0 =\frac{E}{c})$

\begin{eqnarray}
\label{proclu2bis} \begin{array}{l}
 [M_i,M_j] \ = \ i \, \epsilon_{ijk} M_k\, ,
\cr
 [M_i,N_j] \ = \ i\, \epsilon _{ijk}\, N_k\, ,
 \cr
 [N_i,N_j] \ = \ -i\, \epsilon _{ijk} M_k \, ,
\cr
[M_i,P_j] \ =  \i\,  \epsilon_{ijk} P_k\, ,
               \qquad  [M_i, P_0]= 0 \, ,
\cr
[N_i,P_j] \ = \  {i\over 2}\, \delta_{ij}
\left[\kappa c\left(1 - e^{-\frac{E}{\kappa c^2}}\right)
+ \frac{1}{\kappa c} \, \overrightarrow{P}^{\, 2}
\right] -
\frac {i}{\kappa c}\,  P_i P_j\, ,
\cr
[N_i, P_0]= i\, P_i\, , \qquad \qquad  \quad \qquad
[P_\mu, P_\nu] = 0\, ,
\end{array} \end{eqnarray}

b) coalgebra sector

\begin{eqnarray}
\label{proclu3} \Delta(E) & = & E\otimes 1 + 1\otimes E
\cr
 \Delta(\vec{P}) & = &
\vec{P}\otimes 1 + e^{-\frac{E}{\kappa c^2}} \otimes \vec{P}
\cr
\Delta(\vec{M}) & = & \vec{M}\otimes 1 + 1 \otimes \vec{M}
\cr
\Delta(N_i) & = & N_i\otimes 1 + e^{-\frac{E} {\kappa c^2}}\otimes N_i
+ {\frac{1}{\kappa c}}\, \epsilon_{ijk}\, P_j\otimes M_k
\end{eqnarray}

c) antipodes (quantum coinverses)

\begin{eqnarray}
\label{proclu4}
 \begin{array}{l}
S(E) \ = \ -\, E
\\
S(\vec{P}) \ = \ -\, \vec{P}\,  e^\frac{E}{\kappa c^2}
\\
S(\vec{M}) \ = \ -\, \vec{M}
\\
S(N_i) \ = \ - \, e^{\frac{E}{\kappa c^2}}N_i+{\frac{1}{\kappa c}}
\,  \epsilon_{ijk}\, e^{\frac{E}{\kappa c^2}}\, P_j\,  M_k
 \end{array}
\end{eqnarray}

It appears that in the basis given by formulae (\ref{proclu2bis}) the
deformed dispersion relation (\ref{proclu2}) is the deformed mass
Casimir $C^\kappa_2$ for quantum $\kappa$-Poincar\'{e} algebra.

If we introduce the Hopf-algebraic duality relations (see e.g.
 [20]) between
the enveloping algebra $U_\kappa ({\cal P}^{3;1}_{3})$ and
functions $f({\cal P}_\kappa)$ on deformed Poincar\'{e} group
 ${\cal P}_\kappa$
one obtains the full  description of deformed Poincar\'{e}
group. We obtain one-to-one correspondence
\begin{equation}\label{proclu5}
\unitlength0.2cm
\begin{picture}(60,8)
\put(0,4){algebraic sector}
\put(22,4){duality}
\put(40,4){coalgebraic sector}
\put(0,2){of $U_\kappa({\cal P}^{3;1})$}
\put(25,3){\vector(1,0){5}}
\put(25,3){\vector(-1,0){5}}
\put(40,2){of $f({\cal P}_{\kappa})$}
\end{picture}
\end{equation}

\begin{equation}\label{proclu6}
\unitlength0.2cm
\begin{picture}(60,8)
\put(0,4){coalgebraic sector}
\put(22,4){duality}
\put(40,4){algebraic sector}
\put(0,2){of $U_\kappa({\cal P}^{3;1})$}
\put(25,3){\vector(1,0){5}}
\put(25,3){\vector(-1,0){5}}
\put(40,2){of $f({\cal P}_{\kappa})$}
\end{picture}
\end{equation}
In particular because the fourmomenta $P_\mu$ describe Hopf
subalgebra, we obtain

\begin{eqnarray}\label{proclu7}
\unitlength0.2cm
\begin{picture}(60,8)
\put(0,4){$\kappa$-deformed}
\put(22,4){duality}
\put(40,4){$\kappa$-deformed}
\put(0,2){coproducts for ${P}_{\mu}$}
\put(25,3){\vector(1,0){5}}
\put(25,3){\vector(-1,0){5}}
\put(40,2){Minkowski space}
\end{picture}
\end{eqnarray}
where we identified the space-time with the
 the translations sector of the Poincar\'{e}  group. From (\ref{proclu3}) and
(\ref{proclu7}) one gets explicitly the algebra of
$\kappa$-Minkowski space-time, firstly obtained in such a way in 1994
[9,4,5]
\begin{equation}\label{proclu8}
    \left[\widehat{x}_i, \widehat{x}_j
     \right] = 0\, , \qquad
     \left[\widehat{x}_0, \widehat{x}_i
     \right] = \frac{i}{\kappa} \widehat{x}_i\, .
\end{equation}
If we add to the relations  (\ref{proclu8}) the commuting momenta
(see  (\ref{proclu2bis})), we obtain the generators
($\widehat{x}_\mu, \widehat{p}_\mu \equiv P_\mu$) of
noncommutative $\kappa$-deformed relativistic phase space. The
cross commutators between the quantum coordinates and momenta
 $\widehat{p}_{\mu}$ are
derived in a unique way from the quantum
$\kappa$-deformed Poincar\'{e} algebra via the double
cross-product construction, called Heisenberg double [20,10]. One
obtains [4--6]

\begin{eqnarray}\label{proclu9}
\left[ \widehat{x}_0, \widehat{p}_i \right] & = \frac{i}{\kappa} \,
\widehat{p}_i \, ,
\qquad
\left[ \widehat{x}_i, \widehat{p}_j
\right] & = i \hbar \, \delta_{ij}\, ,
\cr \cr
\left[ \widehat{x}_0, \widehat{p}_i \right] & = 0 \, ,
\qquad
\left[ \widehat{x}_0, \widehat{p}_0
\right] & = - i \hbar \, \delta_{ij}\, .
\end{eqnarray}

One can also introduce generalized $\kappa$-deformed phase space,
which is described by the Heisenberg double of complete $\kappa$-deformed
Poincar\'{e} algebra (generators $P_\mu \equiv \widehat{P}_\mu, M_{\mu\nu}$) and
$\kappa$-deformed Poincar\'{e} group (generators $\widehat{x}_\mu,
{\Lambda}_{\mu\nu}$). The  relations (\ref{proclu9}))
have been extended in 1997 [10]
 by the following cross relations:
\begin{eqnarray}\label{proclu10}
    \left[
M_{\mu\nu}, \widehat{x}_\rho
    \right]
    & = & i \, \hbar
    \left(
\eta_{\nu\rho} \widehat{x}_{\mu} - \eta_{\mu \rho}\widehat{x}_{\nu}
    \right)
  + \frac{i}{\kappa}
\left(
\eta_{\nu 0} M_{\mu\rho} - \eta_{\mu 0}M_{\nu\rho}
    \right)
    \cr
 \left[
M_{\mu\nu}, \Lambda_{\rho \tau}
    \right]
    & = & i \, \hbar
    \left(
\eta_{\nu\rho} \Lambda_{\mu\tau} - \eta_{\mu \rho}\Lambda_{\nu\tau}
    \right)
    \cr
\left[
\widehat{p}_{\mu}, \Lambda_{\rho \tau}
    \right]
    & = & 0
\end{eqnarray}

In such a way we obtain a generalized $\kappa$-deformed phase
space with 20 generators
($\widehat{x}_{\mu},{\Lambda}_{\rho \tau},
\widehat{p}_{\nu}, {M}_{\lambda \rho}$) as a
$\kappa$-deformed counterpart of the classical generalized phase space,
  which includes also relativistic spin
degrees of freedom.

Summarizing, the complete information on quantum $\kappa$-deformed
Poincar\'{e} symmetries is given by the Hopf-algebraic form of
$\kappa$-Poincar\'{e} algebra or, equivalently, the
 quantum  $\kappa$-deformed Poincar\'{e} group. Such a structure
extends simultaneously  the notion of classical symmetries in their
infinitesimal (Lie-algebraic) as well as global (Lie group) form.
Further, having $\kappa$-deformed
Poincar\'{e} algebra one can uniquely define via Heisenberg double
construction the $\kappa$-deformed relativistic phase space, in
its standard (8-dimensional) or extended (20-dimensional) form.

The formulation of $\kappa$-deformed Poincar\'{e} algebra can be
given in different bases, related by nonlinear transformations of
the generators. In particular the first formulation of
$\kappa$-Poincar\'{e} algebra has been given in so-called standard
basis [1--3], with deformed boost sector of the Lorentz subalgebra.
The bicrossproduct basis (3--5) with classical
 Lorentz sector can be rewritten in any
nonlinearly related coordinate frame
\begin{equation}\label{proclu11}
    p^{\prime}_{i} = p_i \, f_{\kappa} (\overrightarrow{p}^{\, 2}, p_{0} )
    \, ,
    \qquad
 p^{\prime}_{0} = g_{\kappa}(\overrightarrow{p}^{\, 2}, p_{0} )
 \, ,
\end{equation}
where $f_{\infty}(\overrightarrow{p}, p_0)=1$,
$g_{\infty}(\overrightarrow{p}^{2}, p_0)=p_0$ and we have chosen
the dependence on the length $|\overrightarrow{p}|$ of the
three-momentum in order to preserve in
 all bases  the classical $O(3)$-covariance.
 In particular choosing [7]
\begin{equation}\label{proclu12}
    f_{\kappa}(\overrightarrow{p}^{\, 2}, p_{0}) =
    \frac{A}{\kappa}\,
    e^{\frac{P_0}{\kappa}} \, ,
    \qquad
    g_{\kappa}(\overrightarrow{p}^{\, 2}, p_0) =
    A\left( e^{\frac{P_0}{\kappa}} -1
    - \frac{C_2^{\kappa}\, (\overrightarrow{p}^{\, 2}, p_{0})}{2\kappa^2}\right)
\end{equation}
where  $C^\kappa_2$ is given in by  (2) we obtain $P_\mu^{\prime} = P^{(0)}_\mu$,
(see (1)), i.e. we arrive at  the classical Poincar\'{e} algebra basis.
The standard choice of the formula (12) is obtained if $A=\kappa -
\frac{M^2_0}{2\kappa}$ and $C= \kappa + \frac{M^2_0}{2\kappa}$.
 One gets  the following inverse deformation map: [25,7]
\begin{equation}\label{proclu13}
    P^{(0)}_i = e^{\frac{P_0}{\kappa}} \, P_i \, ,
    \qquad
P^{(0)}_0 = \kappa \, \sinh \frac{P_0}{\kappa} +
\frac{1}{2\kappa} \, e^{\frac{P_0}{\kappa}} \, \overrightarrow{p}^{2}
\end{equation}
where
\begin{equation}\label{proclu13a}
\left(P^{(0)}_0 \right)^2 -
\left(P^{(0)}_l \right)^2 =
M^2_0 =
M^2 \left(
1+ \frac{M^2}{4\kappa^2}
\right)\, ,
\end{equation}
and the formula (15) describes the relation between
the $\kappa$-deformed  rest mass $M$ and its
 classical counterpart $M_0$ in classical Poincar\'{e} basis.
Using (14) we obtain the coproducts of $P^{(0)}_\mu$ given by
 (see [8])
\begin{lit}
\begin{eqnarray}\label{proclu14b}
    \Delta \left(P^{(0)}_0 \right) & = &
    P^{(0)}_0 \otimes K^2  +
    K^{-2} \otimes P^{(0)}_0  + \frac{1}{\kappa} K^{-2}
    \, P^{(0)}_i  \otimes
    P^{(0)}_l
    \\
 \Delta \left(P^{(0)}_i \right) & = &
P^{(0)}_i \otimes K + 1 \otimes P^{(0)}_i
\end{eqnarray}
\end{lit}
where
\begin{equation}\label{proclu14c}
    K  = \kappa^{-\frac{1}{2}}
    \left[
P^{(0)}_0 + \left(\left(P^{(0)}_0 \right)^2 -
\left(P^{(0)}_i \right)^2
+ \kappa^2\right)^{\frac{1}{2}}
    \right]^{\frac{1}{2}}
\end{equation}

The coproduct $\Delta(E)$ of energy $E=P^{(0)}_{0}$
 (we put here $c=1$) describes the Hamiltonian
 of the system composed out of two constituents  [27,10].
 The nonprimitive nature of the coproduct (16a) describes  the
 geometric interaction, which tells us that the system invariant under
 the $\kappa$-Poincar\'{e} symmetry describes geometrically
  interacting
 2-particle system. We obtain from (16a) that [8]
\begin{lit}
 \begin{eqnarray}\label{proclu15}
    E_{1+2} &=& E_1 + E_2 +\frac{1}{\kappa}
    \overrightarrow{P}_1 \, \overrightarrow{P}_2 +
    \frac{1}{2\kappa^2}
[E_2(E^2_2 - \overrightarrow{P}^{\, 2}_2 )
\cr
&&+ \, (E^2_1 +\overrightarrow{P}_1^{\, 2})E_2 -2E_1(\overrightarrow{P}_1\cdot
\overrightarrow{P}_2)]
    +
    O(\frac{1}{\kappa^3})\, ,
\end{eqnarray}
i.e. the terms of order $\frac{1}{\kappa^k}$ for $k\geqq 2$ become
nonsymmetric.
Assuming that $E_1 = \frac{P^2_1}{2\kappa}$, $E_2=\frac{P^2_2}{2\kappa}$
 the formula (18a) can be rewritten as follows:
 \begin{equation}\label{proclu18bis}
    E_{1+2} = \frac{(\overrightarrow{P}_1 +\overrightarrow{P}_2)^2}{2\kappa}
    + O(\frac{1}{\kappa^2}) \, ,
\end{equation}
\end{lit}
It would be very interesting to find a physical interpretation of the
nonlinear terms on rhs of (\ref{proclu15}) as due to some algebraic approximation to
universal
 quantum gravity effects (in such a case one should put $\kappa=M_{\rm pl}$).

It should be added that the transformation of the $\kappa$-Poincar\'{e}
algebra generators from standard to classical basis was studied already in
 1993 by Ma\'{s}lanka      [28].

 In order to introduce the $\kappa$-deformed boost
  transformations of the fourmomentum
 variables one should use the formula (see  [6], formula (2.37))
 \begin{eqnarray}\label{proclu16}
    P_{\mu} (\alpha) & = &
     ad_{\exp \, i \, \alpha_i N_i} ( P_{\mu})
     \cr
     && =  \sum\limits^{\infty}_{k=0}
     \frac{\alpha_{l_1} \ldots \alpha_{l_k}}{k!}
      \, ad_{N_{l_1}}
      \left(
ad_{N_{l_2}} \ldots (ad_{N_{l_k}} (P_k)) \ldots
      \right)
\end{eqnarray}
where  $ad_{a}b \equiv a_{(1)} b S(a_{(2)})$
 denotes quantum adjoint operation [20],
$\Delta(a)= a_{(1)}\otimes a_{(2)}$  (Sweedler notation) and
 $\alpha_i$ denote three boost parameters. The $\kappa$-deformed
 Poincar\'{e} algebra is covariant under the $\kappa$-deformed Lorentz
 transformations (19), and the $\kappa$-deformed  Casimirs remains
 invariant. We see therefore that in the $\kappa$-deformed framework
  we obtain the equivalence of $\kappa$-deformed frames, contrary to the
  framework with modification of mass-shell
   condition  due to broken Poincar\'{e} invariance.

In bicrossproduct basis (3--5), with classical Lorentz subalgebra
 one can derive  the
following relation [6]
\begin{equation}\label{proclu17}
    ad_{N_i} \, P_\mu = [N_i, P_\mu ]
\end{equation}
and the formula (19) takes the form
\begin{equation}\label{proclu18}
    P_{\mu}(\alpha_i) = \exp (i \, \alpha_i \, N_i)
    P_{\mu} \, \exp (-i\, \alpha_i \, N_i) \, ,
\end{equation}
as in the case of classical Poincar\'{e} symmetry. It should be pointed out
 that the formula (21) is also the base for the derivation of $\kappa$-deformed
  boost transformations in DSR1 theory [12]. In quantum $\kappa$-deformed
  framework the formula (21) is  consistent as well with the addition law of the
  momenta described by the coproduct $\Delta(P_\mu)$, (see (4)) i.e.
  one can show that [17]
  \begin{equation}\label{proclu19}
    \left[
    \Delta(P_\mu)
    \right] (\alpha_i) = \exp \left\{
i\, \alpha_i \, \Delta(N_i)
    \right\} \Delta(P_\mu)
    \exp
    \left\{
- i\, \alpha_i \, \Delta(N_i)
    \right\}\, .
\end{equation}
The relation (22) shows how to extend the equivalence of
$\kappa$-deformed Poincar\'{e} frames to multiparticle states.

\section{Doubly Special Relativity (DSR) Theories}

The main idea of DSR theories  is based on
 the physical interpretation of the results of quantum
$\kappa$-deformed framework. It was observed [11--13]  that
in the case of $\kappa$-deformed mass-shell condition (2)
 one can define operationally
the deformation parameter $\kappa$ as the limiting three-momentum. Indeed from
(2)  if $M=0$ it follows that
\begin{equation}\label{proclu19bis}
    \overrightarrow{p}^{\, 2} = c^2 \, \kappa^2 \left(
1-e ^{- \frac{E}{\kappa c}}
    \right)^2
 \mathop{\longrightarrow}_{E\to \infty}
     \, c^2 \, \kappa^2
\end{equation}
i.e. one obtains that the
 maximal value of three-momentum $p\equiv |\overrightarrow{p}|= \kappa c$
   determines the parameter $\kappa$.
It appears therefore that the $\kappa$-deformed Lorentz transformations
(21)
which were explicitly calculated in [12] (see [17] for arbitrary
 boost three-vector)
 leave invariant two parameters:
 the observer-independent  velocity $c$ and maximal momentum $|\overrightarrow{p}|=
\kappa c$, defining the masslike geometric parameter $\kappa$. Identifying
$\kappa=M_p$ one gets the generalized relativity theory with two
fundamental constants $c$ and $M_{p}$, called doubly special relativity
theory.

Unquestionable merit of the DSR reseach is
 stressing  the presence in the $\kappa$-deformed framework of modified
Lorentz    transformation laws between inertial
 observers, and the interpretation of  $\kappa = M_{p}$ as
 the observer-independent
  limit. It should be pointed out however that the equivalence between
  inertial observes in $\kappa$-deformed framework is the basic
  feature of the  whole quantum group approach, and it
   is a build-in property  of
  quantum $\kappa$-Poincar\'{e} framework also in general basis
   (see  [25]). The DSR approach
   borrows however
     only the algebraic part of the quantum
     $\kappa$-Poincar\'{e} framework, and then exposes the property
      that the deformed energy-momentum dispersion
      relation (2) is valid in all $\kappa$-deformed frames.

At present DSR theories determine only the properties of the one-particle sector,
 described by single irreducible representations of $\kappa$-deformed Poincar\'{e}
algebra. In DSR approach  the attitude toward the choice of coproduct is
  ambiguous.
 Usually  (see e.g. [15], where
 the postulates of DSR theories are listed and  discussed)
   there is  presented the symmetric addition law of
   fourmomenta. In such a case the DSR theory is bound to be
   a classical relativistic theory
 with classical linear  coproduct, rewritten in nonlinear basis via
 the  formulae $P_{\mu}=P_{\mu}(P^{(0)})$ inverse to
the relations (14) [7].
  Such an interpretation of DSR1 theories was
 firstly given in [17], where also the nonlinear symmetric addition law
  for energy and momentum
    was derived  [17,18]. Other coalgebra
     extension of DSR1 theory by supplementing the
 addition law
  which is described by the coproduct of quantum $\kappa$-Poincar\'{e}
 algebra (see (4)) is advocated by Kowalski-Glikman et all (see e.g. [16]). This
 version of DSR1 theory is identical with the framework of quantum
 $\kappa$-deformed framework. In particular within
  this approach
  there was investigated the subalgebra of
 $(\widehat{x}_{\mu},{M}_{\mu\nu})$ of $\kappa$-deformed generalized
 phase space (10--11), which is the $D=4$  de-Sitter algebra of
 DSR approach [29].

It should be stressed that from the mathematical point of view the
choice of basis in the  algebraic sector and corresponding choice  of
$\kappa$-deformed mass shell (e.g. given by (2)) is a matter of convention.
In particular for quantum $\kappa$-Poincar\'{e} algebra
 only physical arguments can
select a given
  fourmomentum basis. A good example of such considerations
which provide the definite form of $\kappa$-deformed algebra is the recent
paper by Amelino-Camelia, Smolin and Starodubtcev [30]. In [30] it was
shown that for $D=3$ quantum gravity with cosmological term the
space-time symmetry
 is a $q$-deformed Drinfeld-Jimbo  $D=3$ de-Sitter algebra $SO_q(3,1)$ where $q$
  is proportional to the inverse $dS$ radius.
 In such a way it was shown twelve  years later that the first method of obtaining
  in 1991 the quantum
 $\kappa$-Poincar\'{e} algebra by the contraction of Drinfeld-Jimbo deformation
 of
 ($D=4$ anti-de-Sitter algebra)
  $SO(3,2)$  (see [1]) is based on
  interesting physical ground.

Finally  we would like to recall that besides the DSR1 theories, with
 the range of energies extending to infinity  and limit value
  $|\overrightarrow{p}|=\kappa c$ of three-momenta  (see (23)) we can have
  other two types of DSR theories [26].

  i) with both energy $E$ and three-momentum $|\overrightarrow{p}|$ bounded
   by $\kappa : |\overrightarrow{p}|\leqq \kappa c$ and
    $D\leqq \kappa c^2$. Such theories can be called DSR2 theories [15].
    The first example of DSR2 theory, with symmetric coproduct, i.e.
     equivalent to nonlinear description of classical relativistic theories,
     was presented in [14].

     ii) Remaining class of DSR3 theories is provided by energy bounded
     ($E\leqq \kappa c^2$) and three-momenta unbounded. The example of such
     basis was presented in [26].

     It should be added that one can have obviously the deformed relativistic
     theories with three-momentum and energy unbounded, without the
     interpretation of the parameter $\kappa$ as frame-independent
     limiting value of momentum and/or energy.

  \section{Final Remarks}

  We have an impression that in the framework of DSR theories there is some
  confusion what should be included in the DSR scheme.
  The general attitude of DSR approach is to take only the postulates which are
  physically motivated, and that selects which parts of the quantum
  $\kappa$-Poincar\'{e} framework are incorporated.
  It is however not always so easy:
because of the arguments originating in quantum gravity
the    DSR theories do not
  renounce the noncommutative space-time framework, but
   simultaneously
    the  nonsymmetric coproducts of quantum group
  approach usually  are not accepted.
  Unfortunately you can not have
   both  - or we postulate
  symmetric coproduct and classical space-time, or nonsymmetric
  coproduct and noncommutative Minkowski space. The relation of DSR
  approach to quantum $\kappa$-Poincar\'{e}  framework is therefore
  somewhat schizophrenic. In order to illustrate such a
   statement we recall that
    in some papers (see e.g. [16]) it can be explicitly seen
     that DSR theories
  are described by quantum $\kappa$-Poincar\'{e} framework, but in other
   one (see [31])
     we  find the subtitle about
   ``an illustrative example of
  $\kappa$-Poincar\'{e} algebra which is not admissible in DSR''.

  In conclusion whatever are the disputes  between these two approaches
  one should say that DSR and quantum $\kappa$-Poincar\'{e} points of
  view  have the
  same common aim - to find a physically plausible formalism
  for modification of classical relativistic symmetries.
At present, however,  this goal still has not been fully achieved.

\subsection*{Acknowledgments}
The discussion with G. Amelino-Camelia, J. Kowalski-Glikman,
N. Mavromatos and A. Nowicki
 are gratefully acknowledged.


\begin{thebibliography}{99}
\bibitem{proclu1} J. Lukierski, A. Nowicki, H. Ruegg and V.N. Tolstoy,
Phys. Lett. {\bf B264}, 331 (1991).

\bibitem{proclu2} S. Giller,  P. Kosi\'nski, M. Majewski P. Ma\'{s}lanka
and J. Kunz, Phys. Lett. {\bf B286}, 57 (1992).

\bibitem{proclu3} J. Lukierski, A. Nowicki and H. Ruegg, Phys. Lett.
{\bf B293}, 344 (1992).

\bibitem{proclu4} S. Majid, H. Ruegg, Phys. Lett. {\bf B334}, 348 (1994);
hep-th/9405107.

\bibitem{proclu5} J. Lukierski, H. Ruegg, W. Zakrzewski,
 Ann. Phys.  {\bf 243}, 90 (1995);
 hep-th/9312153.

\bibitem{proclu6} J. Lukierski, H. Ruegg and V.N. Tolstoy, in
{\it "Quantum Groups: Formalism and Applications"}, Proc. XXX-th Karpacz Winter
School, Feb. 1994, ed. by J. Lukierski,
 Z. Popowicz and J. Sobczyk, Polish Scientific Publischers (1995), p. 359.

\bibitem{proclu7} P. Kosi\'{n}ski, J. Lukierski, P. Ma\'{s}lanka,
and J. Sobczyk, Mod. Phys. Lett. {\bf 10A}, 2599 (1995);
hep-th/9412114.

\bibitem{proclu8} J. Lukierski, in ``Quantum Group
Symposium at Group 21'', ed. H.D. Doebner and V.K. Dobrev,
Heron Press, Sofia 1997, p. 173.

\bibitem{proclu9} S. Zakrzewski, J. Phys. {\bf A27}, 2075 (1994).

\bibitem{proclu10} J. Lukierski and A.Nowicki, in ``Quantum
Group Symposium at Group 21'', ed. H.D. Doebner and V.K. Dobrev,
Heron Press, Sofia 1997, p. 186; q-alg/9702003.

\bibitem{proclu11} G. Amelino-Camelia,  Phys. Lett.
{\bf B510}, 255 (2001); hep-th/0012238.

\bibitem{proclu12} R. Bruno, G. Amelino-Camelia and J. Kowalski-Glikman,
Phys. Lett. {\bf B522}, 133 (2001); hep-th/0107039.

\bibitem{proclu13} G. Amelino-Camelia, Int. J. Mod. Phys. {\bf D11}, 35 (2002);
hep-th/0012051.

\bibitem{proclu14} J. Magueijo and L. Smolin, Phys. Rev. Lett.
 {\bf 88}, (2002)190403; hep-th/0112090.

\bibitem{proclu15} G. Amelino-Camelia, D. Benedetti, F. D'Andrea
and A. Bocaccini, Class. Quant. Grav. {\bf 20}, 5353 (2003);
hep-th/0201245.

\bibitem{proclu16} J. Kowalski-Glikman and S.Nowak, Phys. Lett.
{\bf B539}, 126 (2002); hep-th/0203040.

\bibitem{proclu17} J. Lukierski and A.Nowicki, Int. Journ. Mod.
Phys. {\bf A18}, 7 (2003); hep-th/0203065.

\bibitem{proclu18} S. Judes and M. Visser, Phys. Rev.
{\bf D68}, (2003) 045001; gr-qc/0205067.

\bibitem{proclu19} J. Magueijo and L. Smolin, Phys. Rev.
{\bf D67}, (2003) 044017; hep-th/0207085.

\bibitem{proclu20} S. Majid ``Fundations of Quantum Group Theory'', Cambridge University
 Press, 1995.

\bibitem{proclu21} G. Amelino-Camelia, J. Ellis, N.E. Mavromatos  and
 D.V. Nanopoulos, Int. J. Mod. Phys. {\bf A12}, 607 (1997); hep-th/9605211.

\bibitem{proclu22}
 G. Amelino-Camelia, J. Ellis, N.E. Mavromatos, D.V. Nanopoulos and S. Sarkar
Nature {\bf 393}, 763 (1998); astro-ph/9712103.

\bibitem{proclu23} R.J. Protheroe and H. Meyer, Phys. Lett. {\bf B493}, 1(2000).

\bibitem{proclu24} R. Aloisio, P. Blasi, P.L. Ghia and A.F. Grillo, Phys.
Rev. {\bf D62} (2000) 053010.

\bibitem{proclu25}H. Ruegg and V.N. Tolstoy, Lett. Math. Phys. {\bf 32},
 85 (1994); hep-th/9406146.

\bibitem{proclu26} J. Lukierski and A. Nowicki, Proceedings of International
Colloquium
 ``Group 24'', Paris, July 2002, ed. J.P. Gazeau et al. IOP
Publishing House, Bristol, 2003, p. 287; hep-th/0210111.

\bibitem{proclu27} J. Lukierski and P. Stichel, Czech. Journ. Phys.
 { \bf 47}, 55 (1997); hep-th/9606170.
\bibitem{proclu28} P. Ma\'{s}lanka, J. Math. Phys. {\bf 34}, 6025 (1993).
\bibitem{proclu29} J. Kowalski-Glikman, Phys. Lett. {\bf B547}, 291 (2002);
hep-th/0207279.
\bibitem{proclu30} G. Amelino-Camelia, L.Smolin and Starodubtsev,
hep-th/0306134.
\bibitem{proclu31} G. Amelino-Camelia, J. Kowalski-Glikman, G. Mandacini
and A. Procaccini, ``Phenomenology of Doubly Special Relativity'',
gr-qc/0312124.
\end{thebibliography}
\end{document}